\documentclass[twocolumn,showpacs,amsmath,amssymb,pr,superscriptaddress]{revtex4-2}
\usepackage{graphicx}
\usepackage{epsf}
\usepackage{ulem}
\usepackage{color}
\usepackage[unicode=true,colorlinks=true,urlcolor=blue,citecolor=blue]{hyperref}
\usepackage{xcolor}

\newcommand{\ymg}[1]{ #1}
\newcommand{\sus}[1]{ #1}


\begin{document}

\title{Coexistence of two hole phases in high-quality $p$-GaAs/AlGaAs in the vicinity of Landau level filling factors $\nu$=1 and $\nu$=(1/3)}

\author{I.~L.~Drichko}
\affiliation{Ioffe Institute, Russian Academy of Sciences, St. Petersburg, 194021 Russia}
\author{I.~Yu.~Smirnov}
\affiliation{Ioffe Institute, Russian Academy of Sciences, St. Petersburg, 194021 Russia}
\author{A.~V. Suslov}
\affiliation{National High Magnetic Field Laboratory, Tallahassee, FL 32310, USA}
\author{K.~W.~Baldwin}
\affiliation{Department of Electrical Engineering, Princeton University, Princeton, NJ 08544, USA}
\author{L.~N.~Pfeiffer}
\affiliation{Department of Electrical Engineering, Princeton University, Princeton, NJ 08544, USA}
\author{K.~W.~West}
\affiliation{Department of Electrical Engineering, Princeton University, Princeton, NJ 08544, USA}
\author{Y. M. Galperin}
\affiliation{Ioffe Institute, Russian Academy of Sciences, St. Petersburg, 194021 Russia}
  \affiliation{Department of Physics, University of Oslo, PO Box 1048
  Blindern, 0316 Oslo, Norway}

\date{\today}

\begin{abstract}

\ymg{
We focused on the transverse AC magneto-conductance of a high mobility $p$-GaAs/AlGaAs quantum well
 ($p=1.2\times 10^{11}$~cm$^{-2}$)
 \sus{in the vicinity of two values}
 of the Landau level filling factor $\nu$:
$\nu =1$ (integer quantum Hall effect) and $\nu =1/3$ (fractional quantum Hall effect). The complex transverse AC conductance, $\sigma_{xx}^{AC} (\omega)$, was found from simultaneous measurements of attenuation and velocity of surface acoustic waves (SAWs) propagating along the interface between a piezoelectric crystal and the two-dimensional hole system under investigation.}
We analyzed both  the real and imaginary parts of
the hole conductance and compared the similarities and differences between the results for filling factor 1 and filling factor 1/3.
 Both to the left and to the right of these values maxima of a specific shape, "wings", arose in the $\sigma (\nu)$ dependences at those two $\nu$.
 Analysis of the results of our acoustic measurements at different temperatures and \sus{surface acoustic wave} frequencies allowed \ymg{us} to attribute these wings to the formation of collective localized states, \sus{namely the} domains of a pinned Wigner crystal, i.e., a Wigner solid. While the Wigner solid has been observed in 2D hole systems previously, we were able to detect
20 it at the highest hole density and, therefore, the lowest hole-hole interaction reported.

\end{abstract}

\maketitle

\section{Introduction \label{Intro}}

Charge transport through topological materials takes place via surface states. It is topologically protected if bulk electronic states are localized~\cite{PhysRevB.23.5632,AOKI1993951,PhysRevB.25.2185}. This concept \sus{has been frequently} used for the explanation of the plateaus in the integer quantum Hall effect (QHE) in two-dimensional electron gas (2DEG)~\cite{PhysRevLett.45.494}; afterwards it has been generalized for other topological systems. \sus{Accounting for} single-particle (Anderson) localization of electron states in the vicinity of the centers of the integer QHE plateaus turned out to be sufficient for understanding of the integer QHE~\cite{PhysRevB.92.205313}. However, in high-quality structures having low disorder the \sus{accounting for} electron-electron interaction becomes crucially important. As a result, the single-electron states get reorganized into collective ones -- charge density waves~\cite{PhysRevLett.76.499,PhysRevB.54.1853, PhysRevB.54.5006, PhysRevB.55.9326, PhysRevLett.85.5396}. The charge density waves (CDWs) get pinned by disorder, and this way collective localized modes are created.

An example of such modes is the so-called Wigner solid (WS) which is a disordered crystal of electron density forming at the flanks of the integer QHE plateaus~\cite{PhysRevLett.91.016801,PhysRevLett.93.176808,LEWIS2004104}, which is similar to the Wigner crystal (WC) in high magnetic fields~\cite{PhysRevLett.65.633,PhysRevLett.65.2189,Sarma}. Other examples are bubble and stripe phases occurring in a relatively weak magnetic field when several Landau levels (LLs) are occupied. The existence of many competing phases with close free energies leads to rather complicated and diverse behaviors of transport phenomena depending on the electron density, magnetic field, temperature, and other \sus{parameters of a performed experiment}.

\sus{When the Coulomb interaction dominates over the thermal energy of electrons, as well as over their Fermi energy, the theory~\cite{PhysRev.46.1002,JetpLet22.11} predicts implementation of the WS phase.}
 In an \sus{adequately} clean 2D electron system (2DES) placed in a sufficiently high perpendicular magnetic field $B$  all the electrons condense in the lowest LL. If the distance $\hbar \omega_c=\hbar eB/m^*$ to the next LL is greater than the typical Coulomb energy, $E_{\text{C}}=e^2/4\varepsilon_0 \varepsilon \ell_B$, then only the states belonging to the lowest LL are involved \sus{into the interaction}. In this case, the quantum WS will dominate for small values of the filling factor $\nu$, $\nu \le 1/5$~\cite{JetpLet22.11,PhysRevB.30.473,PhysRevB.30.1056,PhysRevLett.111.146804}. Here $\ell_B=\sqrt{\hbar/eB}$ is the magnetic length.

At the same time, the WS competes also with the fractional QHE~\cite{PhysRevLett.48.1559}. As shown in numerous works on 2DES in GaAs quantum wells
\cite{PhysRevLett.60.2765,PhysRevB.38.7881,PhysRevLett.66.3285,PhysRevLett.65.2189,PhysRevLett.66.3285,PhysRevLett.67.1630,PhysRevB.44.8107,PhysRevB.45.13784,PhysRevB.46.7957,Kukushkin_1993,Shayegan_review,PhysRevLett.88.176802,PhysRevLett.89.176802,Chen2006,Tiemann2014,PhysRevLett.117.096601,jang2017sharp,PhysRevLett.122.116601}
the insulating phases are due to forming a WS localized by weak, but inevitable disorder, see, e.g.,~review~\cite{Shayegan_review} and Ref.~\cite{PhysRevLett.125.036601}

Investigation of 2D systems of interacting holes (2DHS) is of special interest~\cite{jang2017sharp,PhysRevLett.82.1744,PhysRevLett.99.236402,PhysRevLett.108.106404,PhysRevLett.68.1188,PhysRevB.46.13639,bayot1994thermopower,PhysRevLett.79.1353,PhysRevB.61.10905,PhysRevLett.92.256804,PhysRevLett.94.226802,PhysRevB.71.035302,PhysRevLett.120.016802,PhysRevB.97.085135}.
The point is that the hole effective mass, $m^*$, in GaAs  is about a half of the free electron mass $m_0$~\cite{zhu2007density}, so it is much larger than the effective mass of an electron in this material ($0.067m_0$). This fact strengthens the role of \sus{the hole-hole} interaction since the corresponding dimensionless parameter, namely the ratio between the inter-particle distance to the effective Bohr radius,
$$r_s^{(2D)} = m^*e^2/\left(4\pi^{3/2}\hbar^2\varepsilon \varepsilon_0\sqrt{p}\right),$$
is proportional to the effective mass $m^*$ of the quasiparticles.
On the other hand, an increase in the effective mass leads to a decrease of the cyclotron frequency $\omega_c \propto 1/m^*$.

Several states of the system can be represented by stability diagrams in several coordinates such as filling factor $\nu$, temperature $T$ and the so-called LL mixing factor, $\kappa$. The latter is defined as the ratio of the typical Coulomb energy, $E_{\text{C}}=e^2/4\pi \varepsilon \varepsilon_0\ell_B$ 
 to the cyclotron energy, $\hbar eB/m^*$,
$$
\kappa \equiv \frac{e^2}{4\pi \varepsilon \varepsilon_0\ell_B}\cdot \frac{m^*}{\hbar eB}\, .$$
If $\kappa$ is large, the LL mixing suppresses fractional QHE in favor of WC~\cite{yoshioka1984effect,PhysRevLett.70.335,PhysRevLett.70.335,PhysRevLett.70.339,PhysRevLett.70.3487,PhysRevLett.71.2777,PhysRevLett.121.116802}. The stability diagram in coordinates $\nu-\kappa$ at $T\to 0$ describes the quantum melting of the WS. Such a diagram is considered in Ref.~\cite{PhysRevLett.121.116802}; it is compared with experimental results for 2DHS obtained in modulation-doped GaAs systems~\cite{PhysRevLett.125.036601}. In that work, the samples with hole densities $p=(2.0-7.9)\times 10^{10}$~cm$^{-2}$ and low-temperature mobility $\mu\simeq 1.5\times 10^6$~cm$^2$/Vs were studied in magnetic field up to 12~T and in the temperature region 40~mK $\div$ 1~K. Those field and temperature domains allowed the authors \sus{of Ref.~\cite{PhysRevLett.125.036601}} to investigate both thermal and quantum melting of the WS.

It turns out that in high-mobility $p$-GaAs/AlGaAs quantum wells a pinned WS can be formed close to
$\nu$=1, 2/5, and 1/3 even at
 $r_s < 37$, see, e.~g., \cite{PhysRevLett.108.106404,app8101909,PhysRevLett.125.036601} and references therein. In the above works samples with low hole density were studied, $p < 10^{11}$~cm$^{-2}$. Note that identification of the WS phase usually \sus{requires auxiliary experiments in addition} to the standard transport measurements. In particular, the authors of~\cite{,PhysRevLett.125.036601} \sus{measured} the screening efficiency~\cite{PhysRevLett.122.116601} while the authors of ~\cite{PhysRevB.97.085135} performed DC $V-I$ measurements.

In the present work we study AC conductance in  high-mobility samples of $p$-GaAs/AlGaAs with  $p=1.2\times 10^{11}$~cm$^{-2}$, i.e., with a density 2-3 times higher than in the samples studied earlier. The complex AC conductance is found by simultaneous measurements of attenuation and velocity of surface acoustic waves (SAWs) propagating in the vicinity of the two-dimensional charge carriers \sus{under study}. Acoustic methods turned out to be useful for studies of pinned Wigner crystal in $n$-GaAs/AlGaAs~\cite{doi:10.1063/1.4975107}. We aim to understand anomalous behaviors of low-temperature AC complex conductance versus magnetic field, temperature, and SAW frequency.

\section{Experimental procedure and results}
\subsection{Procedure}

We have studied \sus{two $p$-GaAs/AlGaAs samples} having a quantum well with width of 17 nm. The high-quality samples were multilayer structures grown on a GaAs (100) substrate. The GaAs single quantum well is positioned between 100-nm undoped spacer layers of AlGaAs, and is symmetrically $\delta$-doped on both sides with carbon. The quantum well is located at the depth 210 nm below the surface of the sample.

Both samples' hole concentration $p$ and the mobility $\mu$, measured at $T=0.3$~K,
as well as $r_s$  and $\kappa$ calculated according to the expressions given  in Sec.~\ref{Intro},
are shown in Table~\ref{t1}.

\begin{table}[h]
\begin{center}
\begin{tabular}{| c| c| c| c| c| c| c| c|}
\hline
 Sample, & $p\times10^{-11}$,  & $\mu \times 10^{-6}$,  & $r_s$ at & $r_s$  at & $\kappa$  at & $r_s$  at  & $\kappa$  at \\
        \#    & cm$^{-2}$           & cm$^{2}$/Vs     & $B$=0     & $\nu$=1  & $\nu$=1 & $\nu$=1/3  & $\nu$=1/3 \\
 \hline
 1& 1.19 & 1.45 & 13 & 18.8 & 8.9  & 32.6 & 5.2    \\
 \hline
 2 & 1.22 & 1.76 & 12.8 & 18.7 & 8.8  & 32.4 & 5.1  \\
 \hline
\end{tabular}
\end{center}
\caption{Parameters of the samples.\label{t1}
}
\end{table}

In our experiments we use the surface acoustic wave (SAW) technique, see Ref.~\cite{PhysRevB.62.7470} and references therein. In this method a SAW propagates along a surface of a piezoelectric (lithium niobate) delay line on either edge of which interdigital transducers are placed to generate and detect the wave. The structure under study is pressed down to the surface of the LiNbO$_3$ substrate by means of springs. The AC electric field accompanying the SAW penetrates into the two-dimensional channel. This AC field induces electrical currents in 2DHS which, in turn, cause Joule losses. The interaction of the SAW electric field with holes in the quantum well results in the SAW attenuation $\Gamma$ and its velocity shift $\Delta V/V$. The general advantage of utilizing this technique is that from measurements of $\Gamma$ and $\Delta V/V$ one can determine the complex AC conductance of the 2D structure, $\sigma^{AC}_{xx}$.   Moreover, the acoustic method does not require electrical contacts, and therefore the results are not affected by them and there is no need in Hall bar configuring. The sample and the piezoelectric crystal are deformation decoupled.

	The experiments were carried out in a dilution refrigerator in the temperature domain 40$\div$300~mK.  We have measured SAW attenuation, $\Gamma$, and variation of its phase velocity, $\Delta V/V$, versus the transverse magnetic field, $B \le 18$~T, at \sus{the SAW basic frequency $f\equiv\omega/2\pi=28$~MHz and its harmonics at frequencies $f$,~MHz:~85, 142, 197, 252 and 306.} In addition, the dependences
of $\Gamma$ and $\Delta V/V$ on the SAW intensity were studied at $T=20$~mK for the same frequencies and magnetic fields. Using the simultaneously measured $\Gamma$ and $\Delta V/V$ we have calculated the components $\sigma_1(\omega)$ and $\sigma_2(\omega)$ entering the complex AC conductance,
\begin{equation}\label{s01}
\sigma^{AC}_{xx}(\omega) \equiv \sigma_1(\omega) - i \sigma_2(\omega)\, .
\end{equation}
The procedure of calculation of $\sigma_{1,2}(\omega)$ from $\Gamma$ and $\Delta V/V$ is described in detail in Ref.~\cite{PhysRevB.62.7470}, see also review~\cite{doi:10.1063/1.4975107}.

\sus{Our findings on magneto-conductivity in samples \#1 and \#2 are very similar and thus everywhere below we present results obtained on sample \#2}.

\subsection{Results and discussion}

\begin{figure}[ht]
\centering
\includegraphics[width=\columnwidth]{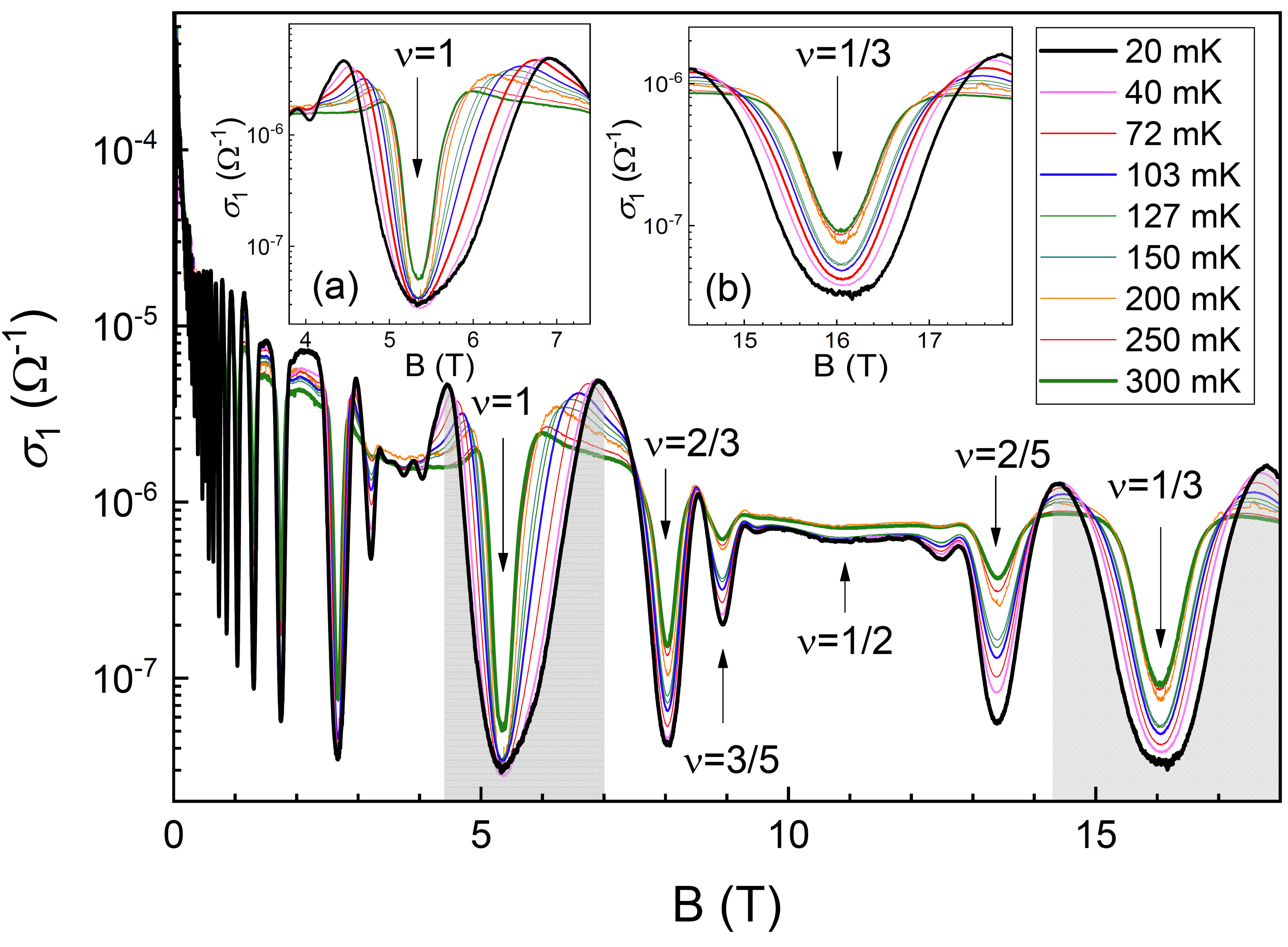}
\caption{Magnetic field dependences of $\sigma_1$ at $f=85$~MHz and different temperatures. Gray regions mark the magnetic field areas analyzed in this article.
Inset: More detailed view of the area  in the vicinity of (a) $\nu=1$ and (b) $\nu=1/3$.
 \label{fig1}}
\end{figure}

Shown in Fig.~\ref{fig1} are the $\sigma_1(B)$-dependences for $f=85$~MHz at \sus{various} temperatures.
One observes a rich oscillation pattern corresponding to the integral and fractional quantum Hall effects. The presence of such a pattern evidences the high quality of the \sus{investigated} structures.
In the following we restrict ourselves by magnetic fields in the vicinity of the filling factors $\nu =1$ and $\nu=1/3$ for which the $\sigma_1(B)$ dependences have \sus{specific} ``wings''~\cite{PhysRevLett.91.016801}. These areas of our interest are marked gray in Fig.~\ref{fig1} and the wings are enlarged in the Fig.~\ref{fig1} insets.

\subsubsection{\sus{Vicinity of $\nu$=1 (B=5.34 T)}}

\begin{figure}[h!]
\centering
\includegraphics[width=\columnwidth]{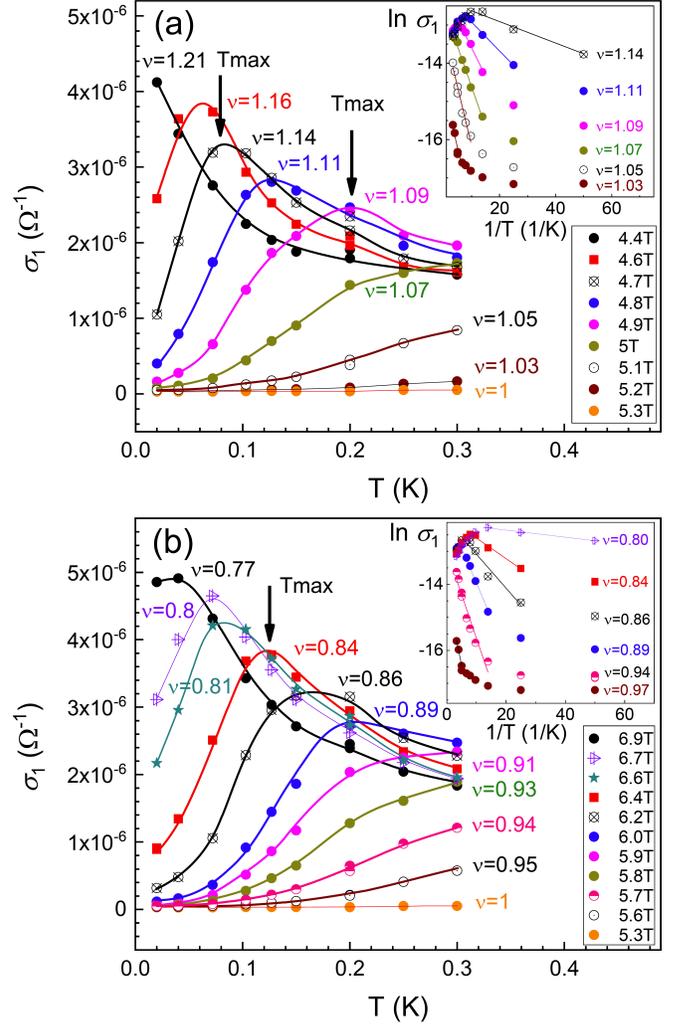}
\caption{Temperature  dependence of  $\sigma_1$ close to $\nu =1$ at $f=85$~MHz and different $\nu$: (a) $\nu \geq 1$, (b) $\nu \leq 1$.  Arrows show the temperatures at which the temperatures dependences change. Lines are drawn as guides to the eye.
Insets: dependences of $\ln \sigma_1$ versus $1/T$ for different $\nu$ and $T < T_{\max}$. Lines are the results of the linear fit.
\label{fig2}}
\end{figure}

Shown in Fig.~\ref{fig2}a are temperature  dependences of $ \sigma_{1}$ for magnetic fields 5.34 $\div$ 4.4~T  ($\nu = 1 \div 1.21$) at the SAW frequency 85~MHz.  Temperature dependences of $ \sigma_{1} $ at $\nu <1$ ($B$=5.34 $\div$ 6.9~T; $\nu = 1 \div 0.77$) are shown in Fig.~\ref{fig2}b and  are qualitatively similar. Thus, we observe that the conductivity behavior is symmetric with respect to $\nu$=1.

\begin{figure*}
\centering
\includegraphics[width=\textwidth]{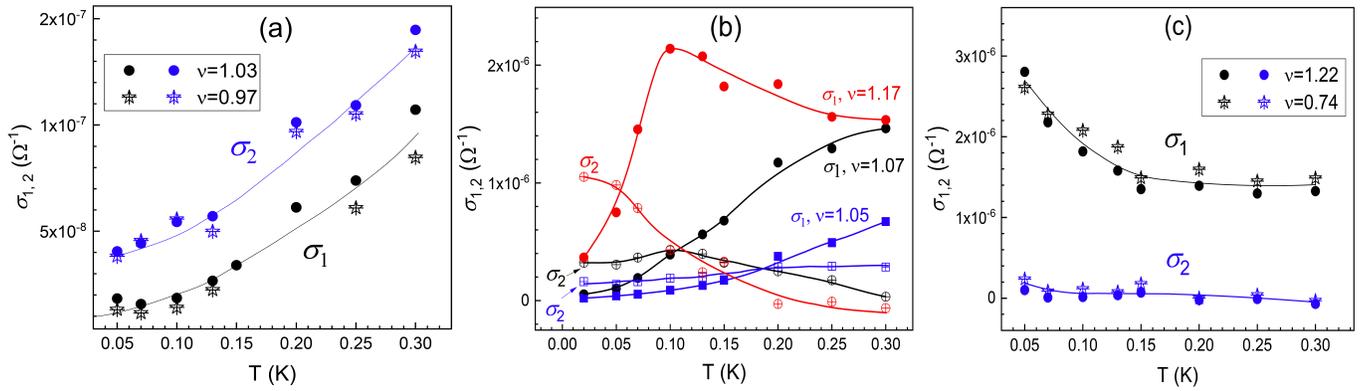}
\caption{Temperature dependences of $\sigma_{1,2}$ for (a) $\nu =1.03$, $\nu = 0.97$, $f=140$~MHz; state A;  (b) $\nu= 1.05, 1.07$ and 1.17, $f=140$~MHz; crossover between states A and B; (c) $\nu =1.19$, $\nu =0.74$,  $f=140$~MHz; state B.
Lines are drawn as guides to the eye.
\label{fig3}}
\end{figure*}

On both sides of $\nu$=1 we can identify 3 regions of the filling factor characterized by different temperature dependences of $\sigma_1$ and by its relation to $\sigma_2$. These regions are presented successively in panels a-c of Figure~\ref{fig3}.

In the first region, see Fig.~\ref{fig3}a, where $|\nu -1| \leq$0.3, i.e., $\nu$ is close to 1, $\sigma_1$ \textit{increases} with temperature. The imaginary part of the AC conductance, $\sigma_2$, is larger than $\sigma_1$ and also increases with temperature at all studied temperatures and frequencies.
Both $\sigma_1$ and $\sigma_2$ are almost independent of frequency.
Let us refer to this state of the system as state A.

In the third region of the filling factor,
see Fig.~\ref{fig3}c, where the deviation $|\nu -1|$ is $\approx$~0.2, i.e. is large, $\sigma_1$ decreases with the temperature increase.
The component
$\sigma_2$ is much less than $\sigma_1$ and remains temperature independent. The difference in the conductivity behavior in the filling factor regions one and three evidences the difference in the conductance mechanisms in these regions. We label the state of the system in region 3 as state B.

In the second (intermediate) region of the filling factor (Fig.~\ref{fig3}b), $0.03\le |\nu -1| \le 0.17$, the temperature dependence of $\sigma_1$ becomes non-monotonic: $\sigma_1$ first increases with temperature, then reaches a maximum and then decreases. Deviation of the filling factor from 1 leads to a decrease of the temperature at which the maximum is reached. Arrows in Fig.~\ref{fig2} correspond to the maximums of $\sigma_1$ at different $\nu$.

Notice that the relation between $\sigma_1$ and $\sigma_2$ changes in the filling factor region 2. According to Fig.~\ref{fig3}b, at low temperatures $\sigma_2 \geq \sigma_1$ which is similar to state A. (Figure~\ref{fig3}b does not include data for $\nu <$ 1, as they look similar to the presented dependences for $\nu >$ 1.) As the temperature increases, $\sigma_2$ decreases and eventually becomes less than $\sigma_1$, as in state B. The larger the deviation of $\nu$ from 1 is, the lower the crossing temperature of the $\sigma_1 (T)$ and $\sigma_2 (T)$ curves is. Due to the observed evolution of both $\sigma_1 (T)$ and $\sigma_2 / \sigma_1$, the region of the filling factor 2 can be considered as a transitional one between the states A and B. A detailed discussion of these states follows.

In the region 1 of the filling factor and in the part of the region 2 where $\sigma_1$ increases with temperature (i.~e., at $T < T_{\max}$)
the temperature dependences of $\sigma_1$ can be considered as having an activation character.
This conclusion is based on the analysis of the $\ln \sigma_1$ versus $1/T$ behavior
presented in the insets in Figs.~\ref{fig2}a and ~\ref{fig2}b.
At relatively high $T$ $\sigma_1$ obeys empirical equation  $\sigma_1 \propto \exp (-\Delta/2k_{\text{B}}T)$.
However, as the temperature decreases the dependence $\ln \sigma_1$ versus $1/T$ flattens that can be interpreted as hopping between the localized states in the impurity band according to the picture of single-electron Anderson localization.
This conclusion is also supported by the absence of the frequency dependences of $\sigma_{1,2}$, as well as by the fact that   $\sigma_2 / \sigma_1 >$~1   at  $T$=40~mK, see Fig.~\ref{fig3}a and \ref{fig3}b.
In Appendix A we describe a theoretical model that qualitatively explains the observed dependences and the $\sigma_2 / \sigma_1$ relation. According to the insets in Fig.\ref{fig2}, $\Delta$ decreases when $|1- \nu|$ rises, i.e., the input of single-particle hopping into the total conductivity reduces.

\begin{figure}[ht]
\centering
\includegraphics[width=.9\columnwidth]{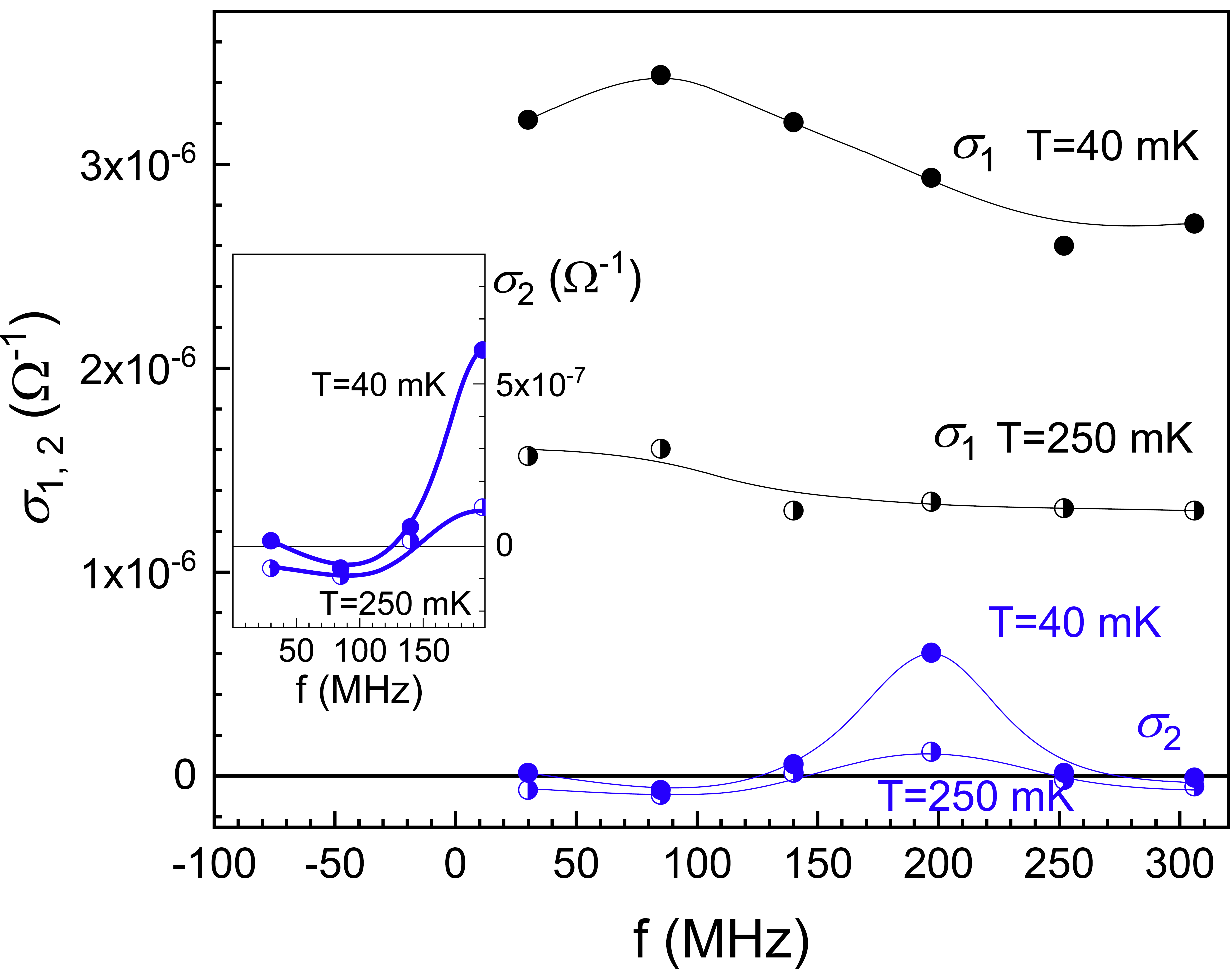}
\caption{Frequency   dependences of $\sigma_1$ and  $\sigma_2$ for $T=40$~mK and $T$=250~mK at $B=4.4$~T ($\nu =1.21$). Inset: Enlarged view of the dependence of $\sigma_2$ on $B$ for low frequencies. Lines are drawn as  guides to the eye.
  \label{fig4}}
\end{figure}

To interpret the state B let us consider the frequency dependences of $\sigma_1$ and $\sigma_2$ for $\nu = 1.21$, Fig.~\ref{fig4}.
The curves in Fig.~\ref{fig4} are less distinct but are similar to the dependences observed on electronic WS in n-GaAs quantum well and calculated in that article following theory~\cite{Fogler-Huse}, see Figs.~4 and 8 in ~\cite{doi:10.1063/1.4975107}, respectively.
The frequency dependences of $\sigma_1$ in Fig.~\ref{fig4} as well as in Figs.~4 and 8 in ~\cite{doi:10.1063/1.4975107}  demonstrate a maximum, whereas $\sigma_2$ at low frequencies is negative, but at some frequency it changes to positive values and shows more or less prominent dip and bump on the flanks of its intersection with the zero line. The positions and magnitudes of all these features depend on the details of the WS state.

Since (i) the curves in Fig.~\ref{fig4} at $T$=40~mK are qualitatively compatible with frequency dependences of $\sigma_{1, 2}$ of a pinned Wigner crystal; (ii) $\sigma_2 <$ 0 at low frequencies; and (iii) $\sigma_1/ |\sigma_2|>$1, we think that the state B corresponds to a collective localization, i. e., to a pinned Wigner solid. Appendix B includes a more detailed explanation based on the theory by M. Fogler and D. Huse~\cite{Fogler-Huse}.

While at $T=40$~mK  the frequency dependences of $\sigma_1$ and $\sigma_2$ are pronounced, at $T=250$~mK both $\sigma_1$ and $\sigma_2$ become essentially independent of frequency. We explain this behavior by WS melting at higher temperatures.

Unfortunately, the estimation of the domain size and the melting temperature is impossible due to the low-grade manifestation of the WS.

As it follows from Fig.~\ref{fig3}b, there is no sharp transition between the states A and B. We \ymg{speculate} that initially islands of WS appear, and as a result the ratio  $\sigma_2 / \sigma_1$ starts decreasing. Note that the ratio $\sigma_2 / \sigma_1$ starts changing as a function of the filling factor \textit{before} one can observe new temperature dependence of $\sigma_1$, i.e., below $T_{\max}$. The  new temperature dependence emerges only when the relative area occupied by the WS becomes large enough. Since the frequency dependences of both components of $\sigma^{AC}_{xx}$ are weak even at 40~mK one can conclude that the WC domains are not fully developed because of weak pinning in a high-quality  \sus{hole} system. Another reason could be the fact that the \sus{WS} domains occupy a relatively small part of the sample.

\subsubsection{Vicinity of $\nu = 1/3 ~(B=16.1$~T) }

Close to $\nu =1/3$ the dependence $\sigma_1 (B)$ also shows characteristic wings, see Fig.~\ref{fig1}.
 \begin{figure}[h!]
\centering
\includegraphics[width=\columnwidth]{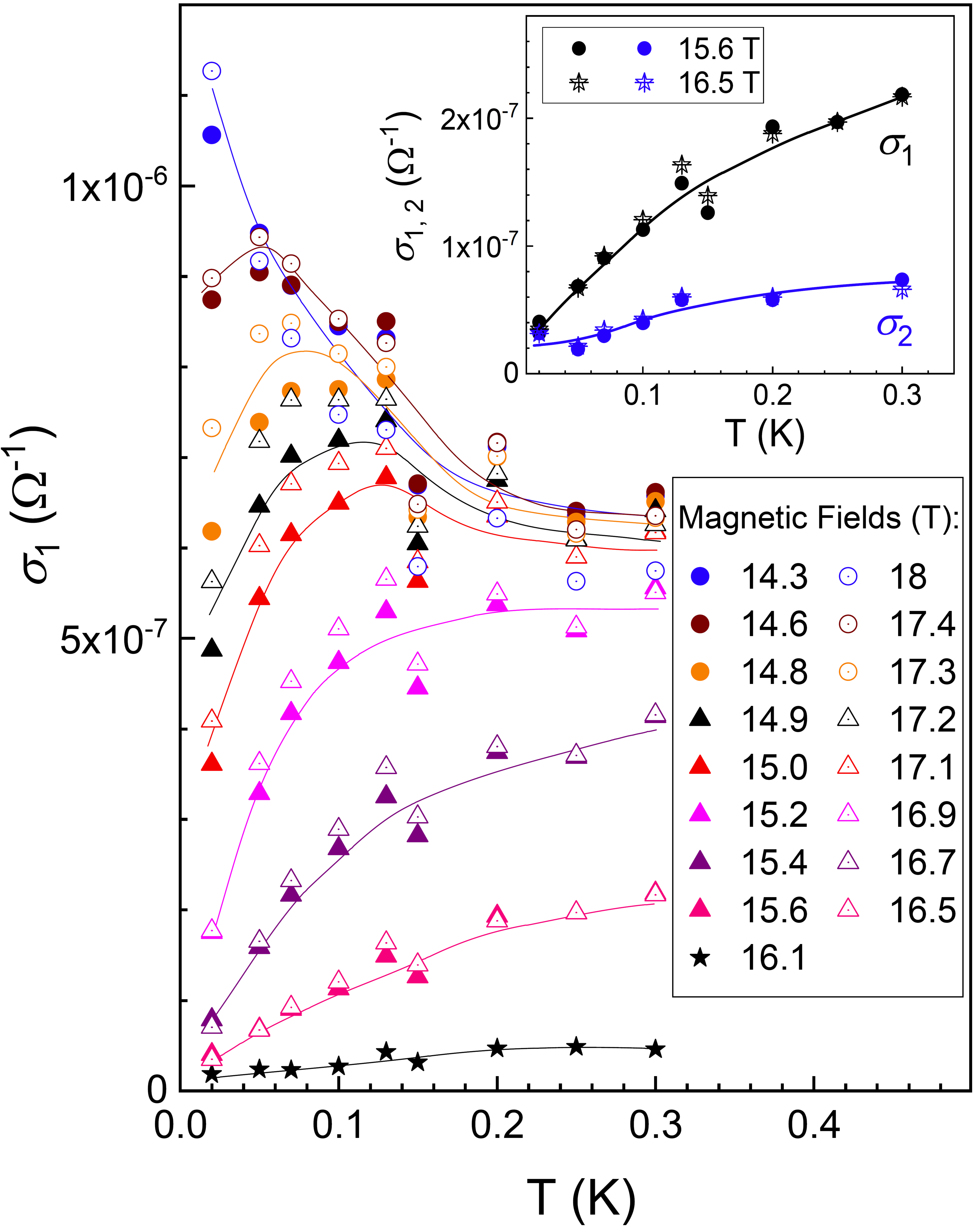}
\caption{Temperature dependences of $\sigma_1$ for \sus{various} magnetic fields close to $\nu=1/3$, $f=140$~MHz. Lines are drawn as  guides to the eye. Inset: Temperature dependences of $\sigma_{1,2}$ for $f=140$~MHz. $\nu$=0.32 ($B=16.5$~T) and $\nu$=0.34 ($B=15.6$~T). Lines are drawn as  guides to the  eye.\label{fig5}}
\end{figure}
Shown in Fig.~\ref{fig5} are the dependences $\sigma_1(T)$ for $\nu = 0.33 \div 0.37$ and for $\nu =0.30 \div 0.33$.
 Figure~\ref{fig5} for the vicinity of $\nu$=1/3 is very similar to Fig.~\ref{fig2} for the vicinity of  $\nu=1$.
  However, as
it follows from the inset of Fig.~\ref{fig5},
 close to $\nu=1/3$ both $\sigma_1$ and $\sigma_2$ increase with temperature, but in contrast with Fig.~\ref{fig3}a in the whole temperature interval $\sigma_1 \gg \sigma_2$.

Notice that the left wing associated with the $\nu=1/3$ oscillation and located in the field below 16.1~T obviously coalesced with the right wing related to $\nu=2/5$. Nevertheless, our analysis shows qualitative agreement between the results obtained on the left and right sides of $\nu=1/3$. Most likely, it is because in the area of our study, $B >14.3$~T, the input from the $\nu=2/5$ state is insignificant for our qualitative analysis, where dependences are guides to the eye rather than fits to theoretical curves.

As seen in Fig.~\ref{fig5} when the magnetic field deviates from 16.1~T ($\nu=1/3$) the component $\sigma_1$ crosses over from increasing to decreasing with temperature.


At $\nu$=0.37 and $\nu$=0.30 the behaviors of $\sigma_1$ and $\sigma_2$ are similar to those shown in Fig.~\ref{fig3}c. The dependences $\sigma_1 (T)$ for different frequencies are similar for all $\nu$ studied in the vicinity of $\nu=1/3$.

We assume that in magnetic fields corresponding to $\nu=0.33 \div 0.353$ (as well as to $\nu=0.313 \div 0.33$) and in the temperature range $T < T_{\max}$ where $\sigma_1$ increases with temperature (see Fig.~\ref{fig5}), the hole state corresponds to the incompressible liquid typical to the fractional quantum Hall effect~\cite{PhysRevLett.125.036601}.

\begin{figure}[ht]
\centering
\includegraphics[width=\columnwidth]{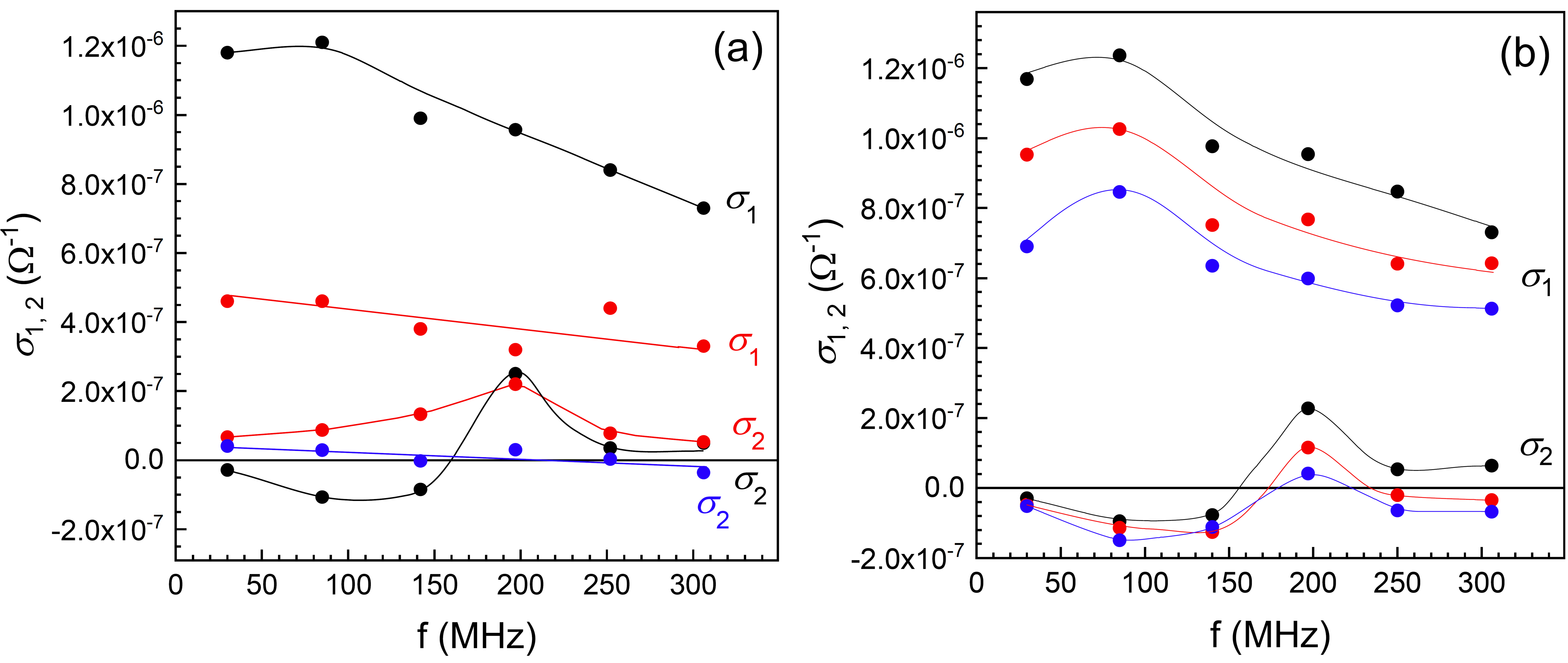}
\caption{(a) Frequency dependences of $\sigma_{1,2}$ at $T=40$~mK for different  $\nu = 0.375$ ($B=14.3$~T) -- black, 0.355 ($B=15.1$~T) -- red, 0.34 ($B=15.7$~T) -- blue;
(b) Frequency dependences of $\sigma_{1,2}$ for  $\nu$=0.375 for different  $T$ (mK): 40 -- black, 130 -- red and  310 -- blue. Lines are drawn as guides to an eye.
 \label{fig7}}
\end{figure}

Shown in Fig.~\ref{fig7}a are the frequency dependences of the $\sigma_1$ and $\sigma_2$ for different $\nu$. The frequency dependence of $\sigma_2$ at $\nu = 0.375$ and $T$=40~mK
is similar to one observed in Fig.~\ref{fig4} for $\nu$= 1.2 and described in Appendix B, and thus leading to the interpretation that the hole system is in a collective localized state.
As $\nu$ approaches the value 1/3,  the frequency dependence of $\sigma_2$ weakens and finally, within our accuracy $\sigma_2$ becomes essentially frequency independent, and thus there is no collective localization at $\nu$=1/3.
 We do not demonstrate the frequency dependence of $\sigma_1$ for \sus{$\nu=0.34$
 because with our accuracy at this filling factor}   $\sigma_1$ is close to $\sigma_2$.
 Shown in Fig.~\ref{fig7}b are frequency dependences of $\sigma_{1,2}$ for $\nu=0.375$ at different temperatures. One observes that when temperature increases the frequency dependence of $\sigma_2$ flattens. However, a trace of frequency dependence can be found even at $T=300$~mK. It seems that domains of WS dissolve gradually as temperature increases, \sus{and the hole-hole interaction depends} on the deviation of $\nu$ from 1/3.

According to the estimates~\cite{PhysRevLett.121.116802}, for a sample with the parameters given in Table 1, $\kappa= E_{\text{C}}/\hbar \omega_c$=5  at $\nu=1/3$.
As it follows from that paper, see Fig.~\ref{fig3} \sus{there},  for $\nu$=1/3 the \sus{hole} system behaves as an incompressible liquid showing fractional QHE. \ymg{However,}  at $\nu$=0.375  the \sus{hole} system is inhomogeneous - consisting of areas of incompressible liquid and WS domains.  Indeed,  see
~\cite{PhysRevLett.121.116802} (Fig.~\ref{fig3}),
the free energies of the QHE and the WS state are very close to each other and  these energies near match. Unfortunately, the phase diagram in the axes $\nu - T_C$ calculated in~\cite{PhysRevLett.125.036601} for $p$-GaAs/AlGaAs with $p = 3.8 \times 10^{10}$~cm$^{-2}$ is not directly applicable to our samples where phase boundaries were not observed. We believe that it is because the \sus{hole} states in our samples are non-uniform and the transitions between the states are not sharp.

\section{Summary and conclusions} \label{Conclus}

Using the acoustic method, we have studied frequency and temperature dependences of the real and imaginary components of the AC conductance of high-mobility \sus{symmetric $p$-GaAs/AlGaAs quantum wells} with $p = 1.2 \times 10^{11}$~cm$^{-2}$. Note that we studied the AC conductance which, in general, did not require the formation of an infinite percolation cluster. The energy absorption took place as a result of phonon-assisted transitions between localized states, either single-hole ones (as in the regions of Anderson localization) or many-hole ones (as patches of WS).


It is shown that at the filling factor  $\nu$=1 and $T=40$~mK the holes are localized in the minimums of the random potential according to the model of the single--electron Anderson localization. Consequently, the transport can be identified as single--hole hopping between localized states.

As deviation from  $\nu=1$ increases, domains of WS appear to be most pronounced at $\nu =1.2$ and 0.78 ($T=40$~mK).

We think that at $\nu =1/3 $ and $T=40$~mK the \sus{hole} state corresponds to the incompressible liquid typical to the fractional QHE. However, as the magnetic field deviates from \sus{the filling factor} $\nu=1/3$ the WS domains also appear. The domains manifest themselves most clearly at  $\nu =0.3$ and  0.375.
To the best of our knowledge WS localization is the only one that leads to negative $\sigma_2$.

As the temperature is increasing, the WS domains are disappearing which can be inferred from the weakening of the frequency dependence of $\sigma_2$, which we interpret as a melting of the WS.

The single-hole Anderson localization and the incompressible liquid state are commonly accepted phases for $\nu=1$ and $\nu=1/3$, respectively, as reported previously, e.g.,~\cite{PhysRevB.23.5632,AOKI1993951,PhysRevB.25.2185,PhysRevB.92.205313,PhysRevLett.125.036601}. Our results included in this article agree with these widely known conclusions. On the flanks of these filling factors three parameters, i.e., (i) the temperature dependences of $\sigma_1$ and $\sigma_2$, (ii) the frequency dependences of $\sigma_1$ and $\sigma_2$, and (iii) the ratio $|\sigma_2|$/$\sigma_1$ drastically differ from those at $\nu=1$ and $\nu=1/3$. Differences in these three parameters are the criteria that manifest the appearance of a hole phase which is different from the one observed at $\nu=1$ and $\nu=1/3$. When the filling factor diverges from $\nu=1$ or $\nu=1/3$, in both cases holes become delocalized. In the trivial case of noninteracting carriers the delocalized holes would form a metallic state thus leading to a positive $\sigma_2$ and to the absence of frequency dependence of $\sigma_1$ and $\sigma_2$ at the frequency range used for our research. Detected in the studied systems, the combination of negative $\sigma_2$ with a large ratio $\sigma_1$/$|\sigma_2| \gg$~1 together with observed frequency dependences of $\sigma_1$ and $\sigma_2$ distinguish the observed phase from the metallic state and make evident the existence of Wigner clusters which are due to hole-hole interaction, as explained in Appendix A. It should be mentioned that the transitions between the observed phases are far from being sharp -- these transitions rather behave as crossovers.

Our analysis of the temperature and frequency dependences of $\sigma^{AC}_{xx}$ leads to a conclusion that single--hole localized states coexist with localized collective states within the area of the wings observed at $\nu=1.2$ and  0.78 at 40~mK. Similarly, the incompressible liquid states characteristic for the QHE coexist with collective localized states in the area where the wings appeared at $\nu=0.3$  and 0.375 at 40~mK. Therefore, the contribution of WC to conductance is masked by the contributions of other mechanisms.

We found that in the nonlinear intensity regime the SAW power affects the AC conductance dependences on the magnetic field in the same way as the temperature does in the linear intensity regime. Namely,  an increase in the acoustic power leads to hole heating, i.~e., to an increase in the hole temperature. Since our analysis of the SAW intensity impact on the conductance does not bring anything new to the interpretation of our experimental results, we do not present it here.

We believe that the main achievement of this paper is that we report the successful registration of a formation of the WS domains at the lowest  hole-hole interaction reported (the highest hole density), happening on the background of single--hole localization near $\nu$=1 and in the FQHE regime near $\nu$=1/3.

\appendix

\section{Region of single-hole localization}
In this section we briefly review the physics behind AC conductance in the case of single-electron localization, mainly following our previous paper~\cite{doi:10.1063/1.4975107} aimed to analyze an $n$-GaAs quantum well.

First, let us recall that in relatively low (medium) mobility  systems showing
only integer quantum Hall effect the behavior of $\sigma_1$ is well
described by the single-electron picture involving electrons trapped by
a random potential in the vicinity of the conductivity minima. According to this picture, at integer $\nu$ the
Fermi level is located in the middle of the distance between the
Landau levels, the electron states are localized by disorder, and
low-temperature DC conductance, $\sigma^{\text{DC}}_{xx}$, is
exponentially small. The AC conductance is determined by electron
hops between nearest potential wells resulting in $\sigma_1(\omega)
\gg \sigma^{\text{DC}}_{xx}$. In such case, the AC response can be
explained by the two-site model, see for a review~\cite{Galperin1991} and references therein.
According to this model, a pair of the electron energy minima is
described as a two-level tunneling system (TLS) with diagonal
splitting $\Delta_d$ and tunneling splitting $\Lambda (r)$, the
interlevel spacing being $E=\sqrt {\Delta^2_d + \Lambda^2(r)}$. At
our frequencies the AC response is due to the relaxation of
the non-equilibrium populations of the minima. The corresponding
relaxation rate can be expressed as, cf. with~\cite{Galperin1991},
\begin{equation}\label{r-time}
\frac{1}{\tau (E,r)}=\frac{1}{\tau_0 (T)}F\left(\frac{E}{kT}\right)\left(\frac{\Lambda(r)}{E}\right)^2\, .
\end{equation}
Here $k$ is the Boltzmann constant, $F$ is the dimensionless
function depending on the relaxation mechanism normalized in order
to get $F(1)=1$. Therefore, the $\tau_0$ has a meaning of the
\textit{minimal} relaxation time for a TLS with the level splitting
$E=kT$. Note that the above expression allows for the correlation created by Coulomb interaction. In particular, each of important pairs contains only one hole~\cite{Galperin1991}. The theory predicts that (with logarithmic accuracy)
\begin{equation}\label{rel2}
\sigma_1(\omega) \propto \min\{\omega, \tau_0^{-1}(T)\}\,, \quad \sigma_2(\omega) \gtrsim \sigma_1(\omega) \, .
\end{equation}

We conclude that at $\nu =1$ the behaviors of $\sigma_{1}$ and
$\sigma_{2}$ in our sample are compatible with the picture of
relaxation absorption of SAW by localized electrons under condition
$\omega \gg \tau_0^{-1}$, see Eq.~(\ref{rel2}). Indeed, estimates
based on Eq.~(\ref{r-time}) show that the main contribution to the
relaxation rate $\tau_0^{-1}$ is due to piezoelectric interaction
between localized electrons and phonons. In this case, see, e.~g.,~\cite{Galperin1991}, $\tau_0^{-1}(T)$ is roughly proportional to $T$ and, respectively, $\sigma_1 \propto T \cdot \omega^0$.

\section{Region of pinned Wigner crystal}

Wigner crystal is a typical state of a clean low-density system of charged fermions  when the interaction energy $E_C$ significantly  exceeds the Fermi energy $E_F$. In a realistic system the WC gets pinned by disorder forming a glass-like state sometimes referred to as a \textit{Wigner glass}.

At small voltage and low temperature the pinned Wigner crystal should  behave as an insulator.
At finite temperature, parts of the Wigner glass experience
correlated hops between different pinned states leading to the
charge transfer. This process is similar to the \textit{creep} of
dislocations~\cite{Ioffe_1987} or pinned vortices in type-II
superconductors~\cite{RevModPhys.66.1125}.

The dynamic response of weakly pinned Wigner crystal at  our frequencies is dominated by the collective
excitations where an inhomogeneously
broadened absorption line (the so-called pinning mode)
appears. It corresponds to collective vibrations
of correlated segments of the Wigner crystal around their
equilibrium positions formed by the random pinning potential. The
mode is centered at some disorder- and magnetic-field-dependent
frequency, $\omega_p$, with its width being determined by a complicated
interplay between various collective excitations in the Wigner
crystal, see, e.~g.,~\cite{Fogler-Huse}. There are modes of two types: transverse (magnetophonons)
and longitudinal (magnetoplasmons). The latter include fluctuations
in electron density. An important point is that pinning modifies
both modes, and the final result depends on the strength and
correlation length, $\xi$, of the random potential.

%

%
%

Based on the theory provided in Ref.~\cite{Fogler-Huse} we found out that the components of the conductivity $\sigma_{xx}^{AC}$ studied in our experiment can be written as:

\begin{eqnarray}
  \sigma_1&=&\sigma_0 u \frac{\omega}{\omega_{p0}} \frac{1+u^2+(\eta
  \omega/\Omega)^2}{[1+u^2+(\eta \omega/\Omega)^2]^2-(2\eta
  \omega/\Omega)^2}, \label{resigma}\\
 \sigma_2&=&\!-\sigma_0 \frac{\omega}{\omega_{p0}} \frac{1+u^2-(\eta
  \omega/\Omega)^2}{[1+u^2+(\eta \omega/\Omega)^2]^2-(2\eta
  \omega/\Omega)^2}  \label{imsigma}
\end{eqnarray}
where $\sigma_0 \equiv e^2p/m^*\omega_{p0}$,  $\Omega \sim \omega_{p0}^2\eta/\omega_c$. $\eta \equiv
\sqrt{\lambda/\beta}$ is the ratio between the shear ($\beta$) and bulk
($\lambda$) elastic moduli of WC,
$\omega_{p0}$ is the pinning frequency at $B=0$,
$\omega_c$ is the cyclotron frequency, $u$ is a constant.

In the studied region of frequencies and when
  $u \gg  1$
\begin{equation}\label{lsigma}
 \sigma_1 \approx \sigma_0\frac{\omega}{ u } \, , \quad
 \sigma_2 =\! \!-\sigma_0\frac{\omega}{u^2}.
\end{equation}
Thus, there is a qualitative agreement between theory ~[\onlinecite{Fogler-Huse}] and our experiment, namely on these two points:  the component $\sigma_2 < 0$ (at frequencies up to ~150 MHz),
and the ratio $\sigma_1/|\sigma_2|$ is  greater
than 1. Inductive response, $\Im \sigma (\omega) < 0$, as well as a large ratio $\sigma_1/|\sigma_2|$ can be considered as a hallmark of pinned Wigner crystal.


\begin{acknowledgments}
The authors are thankful to E.~Green, R. Nowell, and L. Jiao for technical assistance.
The National High Magnetic Field Laboratory is supported by National Science Foundation through NSF/DMR-1644779 and the State of Florida. This research is funded in part by the Gordon and Betty Moore Foundation's EPiQS Initiative, Grant GBMF9615 to L.~N.~Pfeiffer, and by the National Science Foundation MRSEC grant DMR 1420541.
\end{acknowledgments}


%

\end{document}